\documentclass[prb,aps,floatfix,amsmath,amssymb,superscriptaddress]{revtex4}

\usepackage{mathtools}
\usepackage{amsfonts}
\usepackage{amssymb}

\usepackage{graphicx}
\usepackage[all]{xy}

\usepackage{amsthm}

\newtheorem{theorem}{Theorem}[section]
\newtheorem{lemma}[theorem]{Lemma}

\newcommand{\be}{\begin{equation}}
\newcommand{\ee}{\end{equation}}

\begin{document}
\title{Reduced Space-Time and Time Costs Using Dislocation Codes and Arbitrary Ancillas}
\author{M. B. Hastings}
\affiliation{Microsoft Research, Station Q, CNSI Building, University of California, Santa Barbara, CA, 93106}
\affiliation{Quantum Architectures and Computation Group, Microsoft Research, Redmond, WA 98052, USA}

\author{A. Geller}
\affiliation{Quantum Architectures and Computation Group, Microsoft Research, Redmond, WA 98052, USA}

\begin{abstract}
We propose two distinct methods of improving quantum computing protocols based on surface codes.  First,
we analyze the use of dislocations instead of holes to produce logical qubits, potentially reducing spacetime volume required.  
Dislocations\cite{dis2,dis} induce defects which, in many respects, behave like Majorana quasi-particles. We construct circuits to implement these codes and present fault-tolerant measurement methods for these and other defects which may reduce spatial overhead.  One advantage of these codes is that
Hadamard gates take exactly $0$ time to implement.  We numerically study the performance of these codes using a minimum weight and a greedy decoder using finite-size scaling.
Second, we consider
state injection of arbitrary ancillas to produce arbitrary rotations.  This avoids the logarithmic (in precision) overhead in online cost required if $T$ gates are used to synthesize arbitrary rotations.
While this has been considered before\cite{ancilla}, we consider also the parallel performance of this protocol.
Arbitrary ancilla injection leads to a probabilistic protocol in which there is a constant chance of success on each round; we use an
amortized analysis to show that even in a parallel setting this leads to only a constant factor slowdown as opposed to the logarithmic slowdown that might be expected naively.
\end{abstract}
\maketitle

The surface code\cite{surf1,surf2,surf3}, in several different variants, is a promising potential platform for fault-tolerant quantum computation.  Some results indicate that present-day hardware is approaching the threshold for fault-tolerance\cite{nearthresh}.
In these schemes, the idea is first to implement the Clifford group in a topologically protected way by encoding logical qubits within a particular type of stabilizer code; then, some additional operations are added in a way that is not topologically protected.  These extra operations enable universality, and given the ability to implement the Clifford group fault-tolerantly, it is possible to error-correct these additional operations up to a relatively high threshold.

One proposal for achieving universality is to distill magic states which allow implementing $T$ gates by state injection\cite{magic}.  In Ref.~\onlinecite{topt}, it was proposed to implement a quantum computer using alternating rounds of Clifford and $T$ gates.  By teleportation, the Clifford gates are implemented in constant time, regardless of their complexity.
Essentially, one prepares in advance some entangled state using Clifford operations; then, a measurement is used to teleport the logical qubits of the code
while implementing the Clifford operation.  Regardless of the complexity of the Clifford operations when expressed in terms of CNOT and Hadamard gates, the time remain constant, but, if the implementation is done with a surface code, the spacetime volume does increase with increasing complexity.  This scheme is in fact not specific to the surface code, but can be applied to much more general quantum error-correcting codes.
In this scheme, $T$ gates are implemented by state injection; this applies the $T$ gate probabilistically, with half the time applying $T$ and half the time applying $T^\dagger$; however, since $T^2$ is in the Clifford group, this error can be corrected using a Clifford operation from the surface code.  
To implement an arbitrary rotation to accuracy $\delta$ still requires a time logarithmic in $1/\delta$,
even assuming that the magic states to prepare the $T$
gates are impemented perfectly, as the rotation must be compiled into $T$ gates and Clifford operations.

In this paper we consider two distinct topics relevant to this scheme.  First, we consider using dislocations\cite{dis2,dis}, instead of holes, to implement
logical qubits in a toric code.  Second, we consider injecting arbitrary states instead of $T$ gates; this avoids the logarithmic in $1/\delta$ overhead mentioned above.

We provide circuits to implement a disclocation code ``in software", by repeatedly measuring the syndromes of the code.  In such schemes, where classical control is used to perform error-correction, there remain many interesting questions, such as ideal circuits to use and the best error-correction scheme to consider.  Candidate error-correction schemes include minimum weight matching\cite{surf2}, renormalization-group decoders\cite{rgdecode}, and matrix-product decoders\cite{mpdecode}.  We do not consider these questions here, ignoring all issues of the classical
control required.

We then discuss implementing logical operations.  We show that Hadamard can be performed in exactly $0$ time.  We show how to perform CNOT gates by joint measurements.  We discuss several schemes for fault-tolerant measurement, including one that may be useful for other settings.
Part of our analysis is based on the {\it distance} of the code, and on the complexity of certain operations (Hadamard in particular being much simpler, though CNOT is more complicated).
Additionally, we study the statistical properties of error-correction in this code assuming perfect stabilizer measurements; we leave the question of studying imperfect measurements for future simulations.

We then turn to the question of arbitrary ancilla injection in a scheme similar to that mentioned in Ref.~\onlinecite{ancilla}.  Errors in injecting a given ancilla require injecting further ancillas, and so on; this
leads to an interesting question: do the possible delays in implementing a gate on one qubit slow down the other gates by a logarithmically divergent amount?  We show that this does not hold (given some assumptions on the complexity of the intermediate Clifford gates used); this result may be of interest elsewhere in
other schemes such as in forced measurement\cite{motqc} or repeat-until-success\cite{rus}.  All of these schemes, including ours, have the property that
certain gates only succeed with a certain probability on each round.
In forced measurement, for example, if an undesired measurement outcome is achieved, further braids and measurements are done until the desired result is obtained, while in repeat-until-success, additional Clifford operations and measurements are done.
Perhaps a good name for such classes of gates is ``Las Vegas gates", in analogy to Las Vegas algorithms, which are algorithms that are guaranteed to provide the correct answer when they terminate but whose runtime is probabilistic.
We more formally define these gates later and analyze this setting.  We further present an analysis of the number of extra ancillas required under various assumptions on the angles needed.

\section{Dislocations}
\subsection{Review}
The toric and surface codes are local stabilizer codes.
The original toric code\cite{tc} considered a lattice on a torus or other higher genus surface, with the number of logical
qubits depending on the genus of the surface.  In order to implement the system on a physical two-dimensional device, the surface code instead considers a planar system, with the 
logical qubits arising from defects in the lattice.  In this paper, we argue that lattice dislocations, as recently proposed in Ref.~\onlinecite{dis2,dis}, have potential advantages over other choices of defects.  We show how to implement the Clifford group and describe a method for fault tolerant measurement, which also has potential applications to other surface codes.

There are several ways to write a surface code.  We choose, for the model without defects, to consider a square lattice; we color this lattice in checkerboard fashion with light and dark plaquettes.  At each vertex of the lattice there is a single qubit.
There is one stabilizer for each plaquette; for the light plaquettes the stabilizer is the product of $Z_i$ over all the qubits $i$ that border that plaquette while for the dark plaquettes the stabilizer is the product of $X_i$ over all qubits $i$ that border that plaquette.  Here we write $X_i,Z_i$ to denote Pauli $X$ or $Z$ operators on qubit $i$.

On a finite lattice with open boundaries, it is possible to add additional stabilizers supported just on edge of the lattice.  One can either add $X_i X_j$ over a pair of qubits $i,j$ which are on the edge of a light plaquette or $Z_i Z_j$ over a pair of qubits which are on the edge of a dark plaquette.  We refer to the first choice as ``electric" boundary conditions and the second as ``magnetic"; they are sometimes instead referred to as rough or smooth boundary conditions.  Along a given edge of the lattice, one can use exclusively one type of boundary conditions (for example, $X_i X_j$ on {\it all} light plaquettes in that edge) or one can use a mix of boundary conditions.  In the latter case, let us label the qubits on the edge by $i=1,2,3,...$  Then one can have a sequence such as: $X_i X_{i+1}$ on a given light plaquette, followed by $X_{i+2} X_{i+3}$ on the next light plaquette, and so on, up till $X_{i+2m} X_{i+2m+1}$.  Then, one can have $Z_{i+2m+3} Z_{i+2m+4}$ switching to magnetic boundary conditions.  Note that if we compare the total number of stabilizers on an edge which is exclusively electric to the total number of stabilizers on an edge which has $k$ sequences of electric boundary conditions interspersed with $k$ sequences of magnetic boundary conditions, the second choice has $k$ fewer stabilizers.  Indeed, a calculation shows that the second choice of boundary conditions will have $k-1$ extra logical qubits\cite{surf1}, and hence in a sense each switch from electric to magnetic or magnetic to electric adds ``half a qubit".

\begin{figure}
\includegraphics[width=2in]{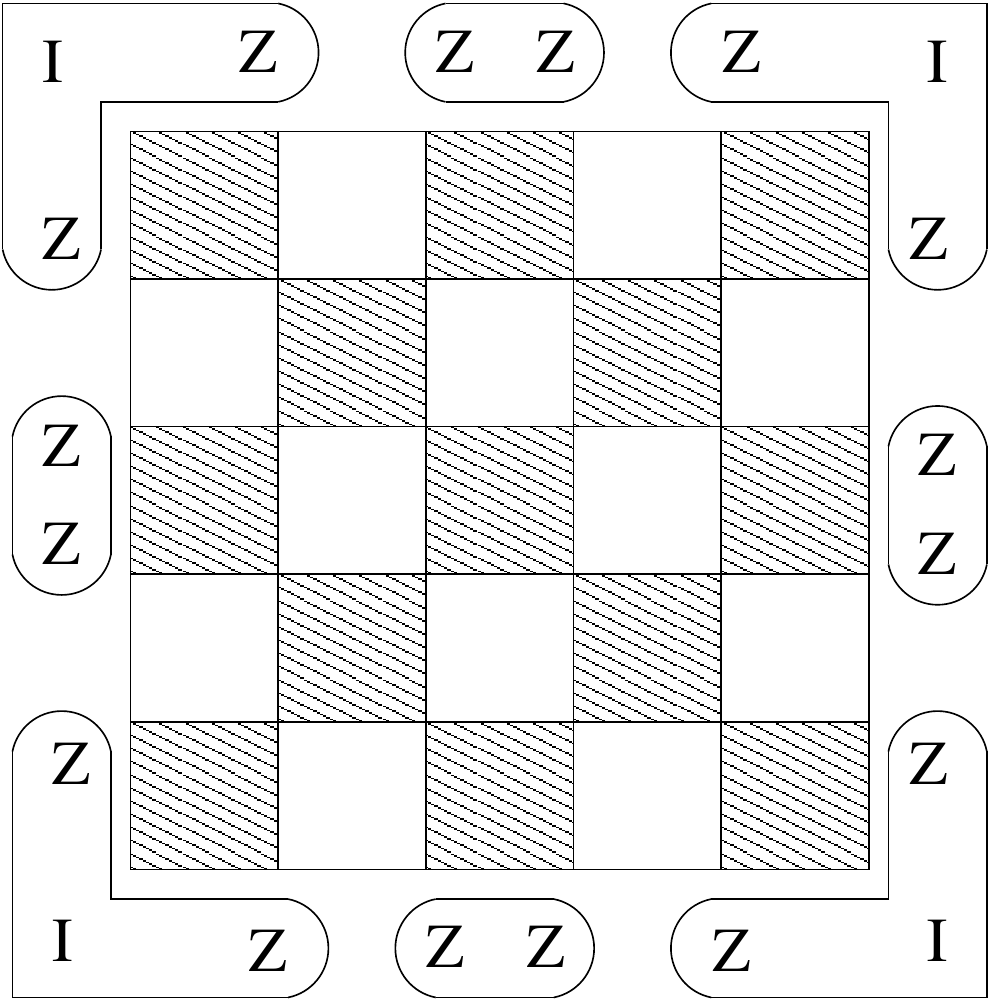}
\caption{Stabilizers for a square with no logical qubits.  Ovals containing $Z$s denote stabilizers with Pauli $Z$ acting on the qubit nearby on the boundary while $I$ denote the identity operator on the qubit.  Additionally, there are four stabilizers not shown, each consisting of a single Pauli $X$ operator acting on one of the four corner qubits.  Thus, there are a total of $37$ stabilizers and $36$ qubits, with one nontrivial product of the stabilizers being equal to the identity.}
\label{fig:square}
\end{figure}

\begin{figure}
\includegraphics[width=2in]{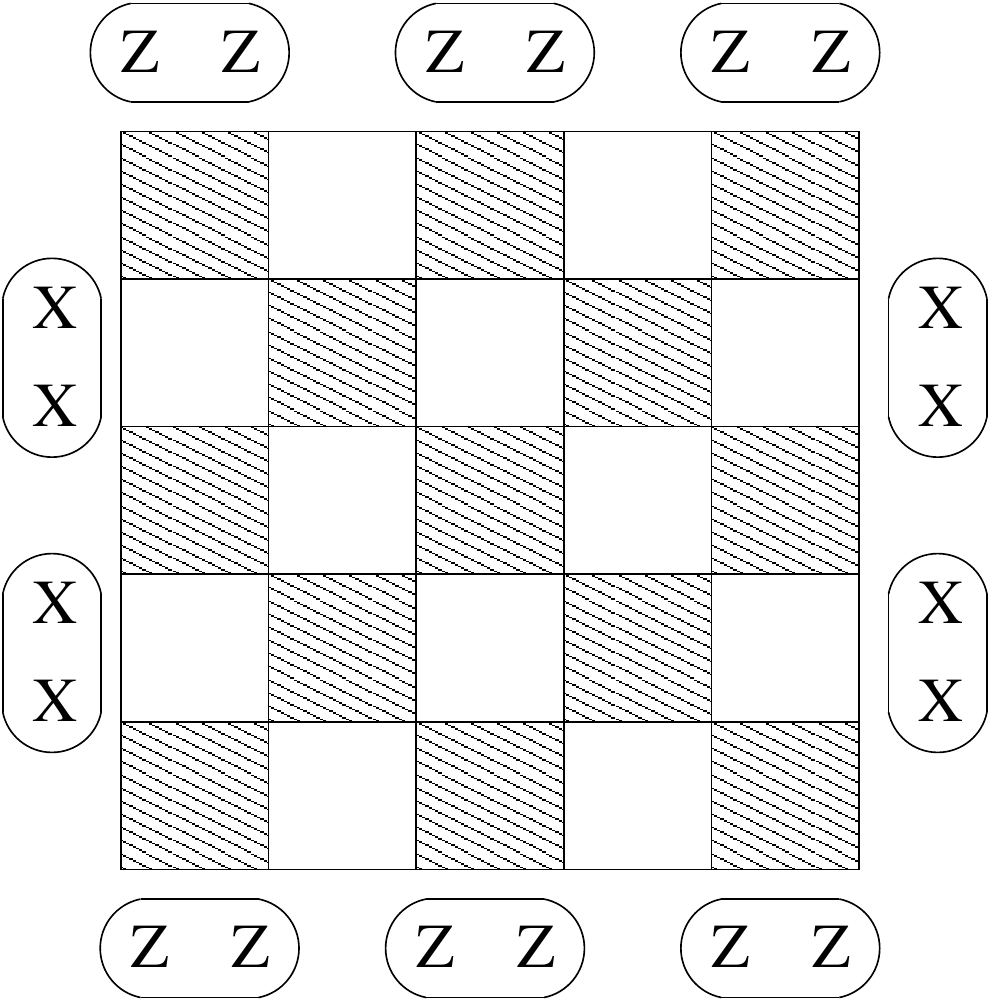}
\caption{Stabilizers for a square with one logical qubit.}
\label{fig:patch}
\end{figure}

Appropriate stabilizers should be added at the corners.  In general, the stabilizers along the boundary should be chosen so that there is no local
operator that commutes with all the stabilizers.
Fig.~\ref{fig:square} shows a choice of stabilizers at the corners so that the boundary conditions are the same everywhere with no logical qubits.
Fig.~\ref{fig:patch} shows a choice of stabilizers giving one logical qubit.

An alternate method to introduce logical qubits is to add holes to the lattice.  The qubits inside some hole are removed (or, at least, they are not used at that point, although they may be physically present), and either electric or magnetic boundary conditions
are chosen for the entire boundary of the hole.  See, for example, Ref.~\onlinecite{measureturnoff}.

Each time a hole is added, this increases the number of logical qubits by one (up to some $O(1)$ corrections, in that, for example, adding a single hole to a code on a sphere does not lead to any logical qubits, but adding $h$ holes leads to $h-1$ logical qubits); however, in order to simplify the implementation of logical gates, usually a {\it pair} of holes are used to represent a single logical qubit.  In the same way, we will also find that it is convenient to increase the number of dislocations required compared to the minimum possible estimate.
These are sometimes referred to as dense and sparse encodings; the main advantage of the sparse encoding for us will be that logical Hadamard becomes trivial to implement.

We note that it is possible to write the same  code in many ways.  By applying a Hadamard transformation to a qubit,
we interchange Pauli $X$ and $Z$ operators on that qubit.  Thus, by a sequence of such transformations, it is possible
to rewrite the surface using the {\it same} operator on every plaquette, regardless of whether it is light or dark.  We can choose this operator to be the product $X_{ne} X_{sw} Z_{nw} Z_{se}$, where $ne, sw, nw, se$ denote the qubits at the northeast, southwest, northwest, and southeast corners of the plaquette.  In an abuse of language, we term these different
ways of writing the same code as a choice of ``gauges", and we refer to this gauge as the ``uniform gauge" to emphasize that
the same operator is used on every plaquette, while we refer to the choice with $Z$ and $X$ operators on light and dark plaquettes as the ``original gauge".
When we later refer to working in one gauge or another, we do not mean that we actually apply the Hadamards to the physical qubits to change the stabilizers; rather, this is done
for notational reasons to simplify certain operators.

\begin{figure}
\includegraphics[width=4in]{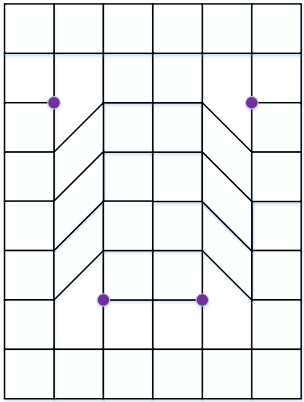}
\caption{Four dislocations in a square lattice.  Qubits marked with a circle have $Y$ Paulis in the stabilizer for the plaquette with $5$ qubits..}
\label{fig:dislat}
\end{figure}

The idea of adding dislocations to a toric code (or more generally any $Z_N$ abelian gauge theory) was introduced in
Refs.~\onlinecite{dis2,dis}.  One simply introduces a lattice dislocation, using this uniform gauge to define the operator on all plaquettes with four qubits in them.  
Fig.~\ref{fig:dislat} shows $4$ dislocations in a square lattice.
Each dislocation
gives one plaquette surrounded by $5$ qubits; on this plaquette we choose a stabilizer which acts on these $5$ qubits, with a Pauli $Y$ on the qubit marked with
a circle.
Dislocations must arise in pairs in order for the boundary of the lattice to have an even number of qubits around the boundary of
the lattice.

If we introduce $k$ pairs of dislocations to the lattice, there are $k-1$ logical qubits (assuming the boundary conditions are chosen
either to be magnetic or electric everywhere, without change from one to another; changing boundary conditions introduces
additional logical qubits).
Borrowing the language of topology, we will say that a {\it cycle} is a product of Pauli operators that exactly commutes with
the stabilizers.  We regard two cycles which are equivalent up to multiplcation by stabilizers as
being in the same homology class.  A trivial cycle is one which is in the same homology class as the identity operator.
In an annulus surrounding an odd number of dislocations it is impossible to consistently color plaquettes as light and dark so that plaquettes of the same color do not neighbor each other, but around a pair of dislocations such a coloring is possible.
Returning to the uniform gauge in an annulus around a dislocation pair, there are cycles which measure either the electric or magnetic charge inside that annulus; however, these two charges must be equal to each other, so that the annulus contains either the identity particle or the $em$ particle, where the $em$ particle is a product of an electric and a magnetic particle.

\begin{figure}
\includegraphics[width=2in]{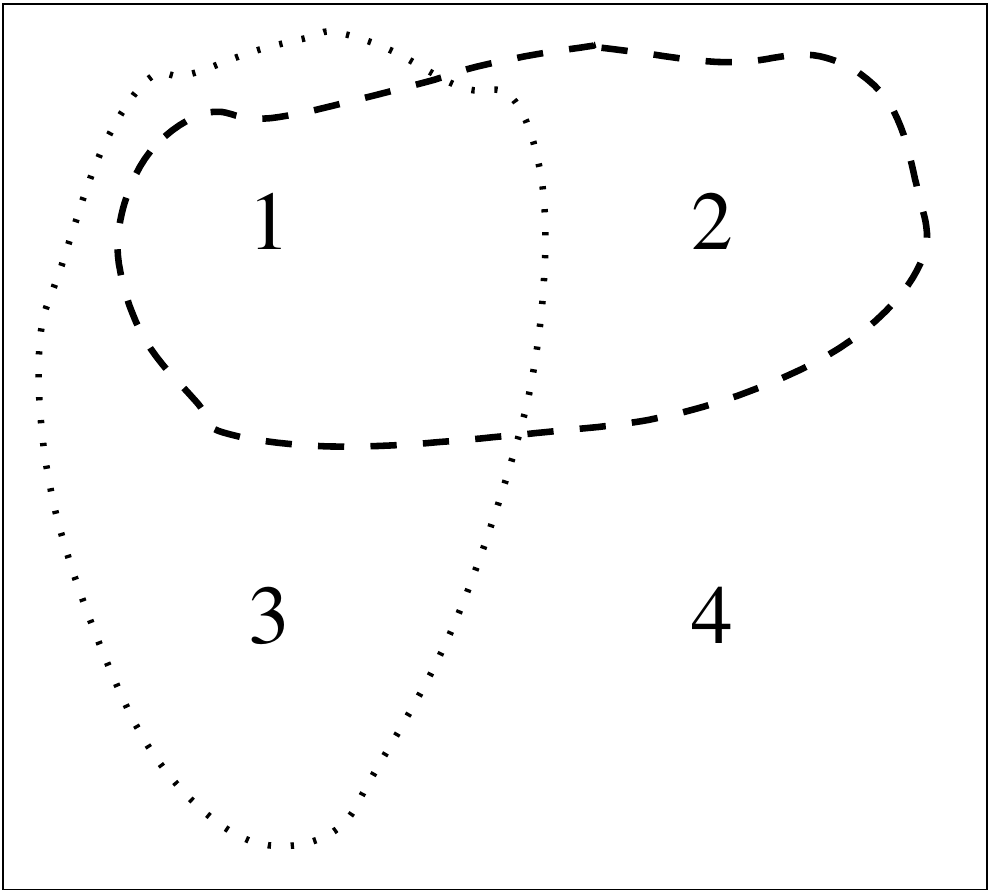}
\caption{Sketch of four dislocations, labelled $1,2,3,4$ in a larger system.  The physical qubits are not shown, merely the location of the dislocations.  The dashed and dotted lines represent two possible logical operators; going to the original gauge around the pair enclosed by a given loop, these would be products of either $X$ or $Z$ around the loop, while in the uniform gauge they
are alternating between $X$ and $Z$ operators along the loop.  These operators {\it anti}-commute.}
\label{fig:4dis}
\end{figure}

Now consider $4$ dislocations as shown in Fig.~\ref{fig:4dis}.  Considering the two different nontrivial cycles shown, we find that they anticommute and can thus serve as logical $X$ and $Z$ operators using $4$ dislocations to encode a qubit.  
Indeed, as shown in Ref.~\onlinecite{dis}, each dislocation can be regarded as representing a Majorana fermion $\gamma_a$, where $a$ labels the dislocation.  The correct
commutation relations for the nontrivial cycles are obtained when the logical operator corresponding to the cycle is the product
of $\gamma_a$ over all $a$ inside the cycle (there are no nontrivial cycles enclosing an odd number of dislocations).
Alternate constructions of Majoranas appear in Refs.~\onlinecite{surf1,bc}.

We now analyze the performance of these codes using dislocations.  Some of our comparisons to other choices
of defects will be based on analyzing either the distance of the code or the time complexity of logical operators.
At the end of the section, we numerically analyze the probability of logical error at non-zero physical error rate; the threshold for the dislocation code
seems to be the same as other surface codes to within finite-size error, as expected since all correspond to the same phase transition
in a random-bond Ising model or random-plaquette gauge model\cite{surf2}.

We begin by giving circuits to perform syndrome measurements in the dislocation code and discuss error correction.  We then compare the performance
of the code as a quantum memory in terms of dislocation density.  Then, we discuss operations, including a fault tolerant
method for measurements.

\subsection{Circuits and Implementation}
\label{imp}
In order to physically implement the dislocation code, we need a means of measuring stabilizers.  For the usual surface codes, it is possible to measure stabilizers using an operation that consists of a total of $8$ rounds as reviewed in Ref.~\onlinecite{measureturnoff}, using the original gauge.  One adds an additional ancilla qubit for each stabilizer.
We will refer to the qubits in the code as ``data qubits" to differentiate them from these ancilla qubits.  Physically the ancilla qubits are centered in the middle of a plaquette so that the four data qubits in the corresponding stabilizer are its neighbors.
Then, in the first round, all ancilla qubits are initialized to $Z=+1$.  In the second round, Hadamards are executed on the ancilla corresponding to stabilizers which are products of four $X$ operators; i.e., these are those in the dark plaquettes.  On the third through sixth round, CNOT gates are executed between data qubits and ancilla qubits.  On the seventh round, the Hadamards are again executed on ancillas corresponding to products of four $X$ operators.  Finally, the ancillas are measured to read out the stabiizers.

The CNOT gates are executed in a particular pattern.  On each of the four rounds from the third through the sixth, each ancilla qubit participates in a CNOT with one of the four neighboring data qubits.  These particular neighboring qubits are executed in a ``$Z$-shaped pattern" such as northwest, northeast, southwest, southeast from the third through the sixth round, respectively.
For the ancillas in the light plaquettes, the CNOT uses the ancilla as a target and the data qubit as a source, while for the ancillas in the dark plaquettes, the orientation is reversed.

This pattern is chosen so that in each of these four round, every data qubit particpaptes in exactly one CNOT, and further, the sequence of measurements is chosen to maintain the correct commutation relations between stabilizers.  That is, consider a pair of stabilizers, $1,2$ which overlap on a pair of qubits $s,t$.  Let $a_1,a_2$ be the ancilla qubits corresponding to the stabilizers.
Then, either $a_1$ interacts with $s$ before $a_2$ interacts with $s$ and $a_1$ interacts with $t$ before $a_2$ interacts with $t$, or else $a_2$ interacts with $s$ before $a_1$ interacts with $s$ and $a_2$ interacts with $t$ before $a_1$ interacts with $t$.

At first, it seems difficult to implement these measurements using the same number of rounds for the dislocation code, because we cannot use the original gauge.  However, in fact there is no obstacle.  It is possible to again present an $8$ round protocol using the uniform gauge.  We begin by describing how to measure all stabilizers {\it except} the stabilizers involving five qubits.  Compared to the case above, all stabilizers are treated equally, rather than having different operations for light and dark plaquettes.  Again we introduce one ancilla per stabilizer, and
again on the first round the ancilla qubits are initialized to $Z=+1$.  Then, on  the second round, we do a CNOT from the data qubit on the northwest corner to the given ancilla.  On the third round, we executed a Hadamard on all ancilla qubits.
On the fourth round, we do a CNOT from ancilla to the data qubit on the northeast corner.  On the fifth round, we do a CNOT from
the ancilla to the data qubit on the southwest corner.  On the sixth rough, we again execute a Hadamard on all ancillas.  On the seventh round, we execute a CNOT from the data qubit on the southeast corner to the ancilla.  On the eigth round, we measure the ancilla.
By using the same ``$Z$-shaped pattern" for measurements, we again maintain the correct commutation relations.

This procedure does not allow us to measure the stabilizers at dislocations which involve {\it five} data qubits, and so these stabilizers cannot be measured in every round.  However, it is possible to measure them less frequently, by also measuring some of the neighboring stabilizers less frequently, so that all stabilizers far from the dislocation are measured in every round and those near are measured in a constant fraction of the rounds.

Error correction using minimum weight matching is in principle the same as other surface codes.  Given every pair of stabilizers,
one first must compute the minimum weight pattern of errors that leads to defects on that pair of stabilizers.  Unlike codes without
dislocations, where the only pairs that can be connected are those where both stabilizers correspond to plaquettes of the same color, now it is possible to connect arbitrary pairs of stabilizers.  Then, given these weights for pairs of plaquettes,
one can compute a minimum weight matching given a set of syndrome measurements.
Note that connecting certain pairs of stabilizer may require a chain that encircles a dislocation in some particular way; we have implemented a computer code that finds such error chains for arbitrary dislocation lattices.

In fact, this error correction does not require knowledge of the stabilizer near the dislocation that acts on five qubits.  For the value of this stabilizer to change, it is necessary for the value of some other stabilizer to change also.  Hence, one might try error correction without explicitly measuring these stabilizers.

\subsection{Comparison as Memory}
We now consider the density of logical qubits when the code is used as a quantum memory.  This next discussion of distances assumes perfect stabilizer measurements.
In the original toric code on a torus, there are $2$ logical qubits.  On a code of size $L$-by-$L$, the distance is $d=L$.  Hence, to achieve distance $d$ means a ratio of physical to logical qubits equal to $d^2/2$.
Consider a dislocation code, with a square lattice of dislocations, with dislocations separated from each other by distance $L$ (we assume that the square lattice is much larger than $L$).
Then, the distance of the code is equal to $2L+O(1)$ (note that the shortest nontrivial cycle must encircle a pair
of dislocations).  Thus, using $4$ dislocations to represent a logical qubit, the ratio of physical to logical qubits is
equal to $d^2-O(d)$.
This can be improved using only $2$ dislocations for a logical qubit giving a ratio $d^2/2-O(d)$; if our goal were simply to obtain a dense memory we could do this, but using $4$ dislocations simplifies the implementation of gates, in particular the Hadamard.

Now consider a surface code using a square patch of size $L$-by-$L$ with electric boundary conditions on the north and south sides and magnetic boundary conditions on the east and west sides.  In this case, we have $1$ logical qubit and again arrive
at the ratio of physical to logical qubits of $d^2$.
Thus, this choice matches that of the dislocation code; however, as we have noted the dislocation code can be made denser by up to a factor of $2$ using a denser encoding, and further the dislocation code will have advantages when performing operations.
We note that this result for the ratio of physical to logical qubits occurs only using the patch oriented as we have shown, with
the plaquettes parallel to the boundaries of the square.  Often, one instead writes the toric code with the degrees of freedom
on the edges of a square lattice; in this case, if the edges of the patch are parallel to the edges of the lattice, we find that
the stabilizers are at a $45$-degree angle to the edges of the square.  Then, the ratio of physical to logical
qubits is $2d^2-O(d)$, which is worse; as an example of this rotated geometry, see Fig.~3 of Ref.~\onlinecite{measureturnoff}.

Finally, we can consider a surface code with holes in it, with either electric or magnetic boundary conditions on each hole.
Conside a square lattice of holes, with each hole having boundaries of size $l$ and the separation between hole centers equal to
$L$.  The distance is equal to ${\rm min}(4l,L-l)+O(1)$.  The optimum is at $l=L/5$, giving $d=4L/5+O(1)$.
Then, using a pair of holes for a logical qubit, the ratio of physical to logical qubits is equal to
$2*(L^2-l^2)=2*(24/25) L^2=3d^2$ up to $O(d)$ corrections, giving a significantly worse ratio.  In fact, the ratio is in practice even worse than that, as in practice the physical qubits inside the holes will also be
present, even if they are not being used.
The scheme of Ref.~\onlinecite{topt}, and other schemes using the surface code, require the use of holes to perform CNOTs, so it seems not to be possible to
use only patches.

Thus, by reviewing these different possibilities, the dislocation code offers asymptotically the highest possible density of any planar code in its dense encoding, and in its sparse encoding it has asymptotically the same density as the optimal patches, while also allowing logical operations.
This motivates further consideration of the code.

It is worth also remarking that it is very natural to consider the dislocation code using {\it three} dislocations per qubit; note that the logical operators in Fig. \ref{fig:4dis} only ever encircle some subset of the first three dislocations and do not use the fourth.  In this case, it is necessary to have an even number of qubits to have an even total number of dislocations.
This increases the density by a factor of $4/3$, making it exceed the optimum density for patches and in fact all the logical operations can still be performed as readily using three dislocations per qubit, rather than four.

We have only considered a square lattice of dislocations above.  Using a triangular lattice of dislocations (or holes) inside the square lattice of the physical code offers no improvement in distance.

\subsection{Logical Operations and Fault Tolerant Measurements}
We now describe how to perform logical operations from the Clifford group in the dislocation code, using a sparse encoding with either
three or four dislocations per qubit.
We will describe how to implement Hadamard gates and CNOT gates.

Consider a given quadruplet of dislocations, labeled $1,2,3,4$.  We use a cycle encircling $\gamma_1 \gamma_2$ to 
measure logical $Z$ for that qubit.  We use a cycle encircling $\gamma_1 \gamma_3$ to measure logical $X$.  Simply making this measurement around a cycle would not be fault tolerant; later we explain how to make it fault tolerant.
We implement the Hadamard gates without actually implementing them: each time we implement a Hadamard gate
on that qubit, we simply interchange whether we will use $\gamma_1 \gamma_2$ or $\gamma_1 \gamma_3$ to measure logical $Z$ on that qubit whenever a measurement is performed in future.
Performing logical $Z$ or logical $X$ operations can be performed by executing the appropriate products of Pauli operators.
Initialization of a qubit in a desired $Z=+1$ state can be performed by measuring $Z$ and then, if the measurement is $-1$ performing $X$ (or simply modifying the subsequent operations to take into account that fact that the qubit started in a $-1$ state).
However, it is easier to perform the logical $X$, $Y$, or $Z$ operations without actually doing any implementation: these unitaries simply change the sign of the expectation value of certain operators.
An $S$ gate can be performed similarly; it interchanges the expectation value of $Y$ and $X$ operators (up to signs) so that future $X$ measurements get replaced with $Y$ measurements and vice-versa.  That is, any single qubit Clifford operation does not need to actually be executed as a circuit, but simply modifies which measurements will be performed in the future.  For each qubit, one must keep track of how the Pauli frame has changed; this can be done classically.

The key to implementing a CNOT is that we have the ability to implement a joint parity measurement $Z_a Z_b$ on two different
logical qubits $a,b$ by measuring the appropriate cycle.
Using the ability to perform joint parity measurements and an ancilla, we can perform a CNOT; see for example Ref.~\onlinecite{cnotjointmeasure}.  Thus, all Clifford operations are
performed simply by measurements.
Note that all these measurements (after arbitrary sequences of Hadamards and $S$ gates) use only three of the dislocations, rather than four, so if the number of qubits is even one can equally well use three dislocations per qubit which slightly increases the density at the same distance.

Let us illustrate this procedure in a specific example (the general case is given in the next paragraph and may be simpler to follow than this specific example).  For example, suppose we prepare $2$ qubits in a state $|0 0 \rangle$, and then apply Hadamards on each qubit, an $S$ gate on qubit $2$, followed by a CNOT gate from qubit $1$ to qubit $2$.  The initialization in state $|0 0\rangle$ is done as described above.  We write the sequence of operations as a product of unitaries ${\rm CNOT}_{1 ,2} S_2 H_1 H_2$.  We need, as in the scheme of Ref.~\onlinecite{cnotjointmeasure}, one extra ancilla.  The CNOT gate uses also a sequence of Clifford operations (in this case, Hadamards), as well as joint measurements $Z_1 Z_a$ and $Z_2 Z_a$.  The first joint measurement done in that protocol is a measurement of $Z_1 Z_2$ after applying $H_a$.  Thus, up to the point of the first joint measurement, we must implement $H_a S_2 H_1 H_2$.  The $H_1$ interchanges $Z$ and $X$ on qubit $1$, while the $H_a$ interchanges $Z$ and $X$ on the ancilla.  So, instead of measuring $Z_1 Z_a$ we measure $H_1 H_a Z_1 Z_1 H_1 H_a=X_1 X_a$.  The next joint measurement is between the ancilla and qubit $2$ after applying a further Hadmard to the ancilla and a further Hadamard to qubit $2$.  Thus, instead of measuring $Z_a Z_2$ we want to measure (combining all the Clifford operations done up to that point): $U^\dagger Z_a Z_2 U$, where $U=H_a H_a H_2 S_2 H_2$.  So, we must instead measure $Z_a Y_2$ (the $H_2$ gate conjugates $Z_2$ to $X_2$, then the $S_2$ gate conjugates $X_2$ to $Y_2$).

In general, our procedure is that each CNOT gate gets expanded into some product of joint $Z$ measurements and single qubit Clifford operations.  Then, whenever we need to implement some long sequence such as $$...M(a'',b'') W M(a',b') V M(a,b) U$$ where $M(a,b)$ denotes a joint $Z$ measurement on qubits $a,b$ and $U,V,W,...$ are some products of single qubit Cliffords, we instead implement
$$...   \Bigl(U^\dagger V^\dagger W^\dagger M(a'',b'') W V U\Bigr) \Bigl(U^\dagger V^\dagger M(a',b') V  U\Bigr) \Bigl(U^\dagger M(a,b) U\Bigr),$$
which leads to exact the same probabilities of measurement outcomes.
Since the $U,V,W,...$ are all products of single qubit Cliffords, each measurement such as $ \Bigl(U^\dagger V^\dagger W^\dagger M(a'',b'') W V U\Bigr)$ is simply a measurement of
the product of some Pauli operator on qubit $a''$ and some other Pauli operator on qubit $b''$.
We can measure any product of Paulis (such as $X_a Y_b$) such as easily as measuring $Z_a Z_b$ by choosing the appropriate cycle (i.e., in this case, by encircling $\gamma_1 \gamma_3$ in the triplet encoding qubit $a$ and $\gamma_2 \gamma_3$ in the triplet encoding qubit $b$).

There are two basic techniques described in the literature to perform fault tolerant measurements of logical operators.  The operator that one desires to measure is, for example, the product of $Z$s around some loop.  The first method, described in Ref.~\onlinecite{basismeasure}, is to measure all qubits in the $Z$ basis in an annulus that includes that loop.  Using classical processing, one can then make the probability of an
error in the logical measurement exponentially small in the width of the annulus.  This procedure can readily be adopted to the uniform gauge; in this case, we measure each qubit in either the $Z$ or $X$ basis depending upon its location relative to the desired logical operator.

\begin{figure}
\includegraphics[width=2in]{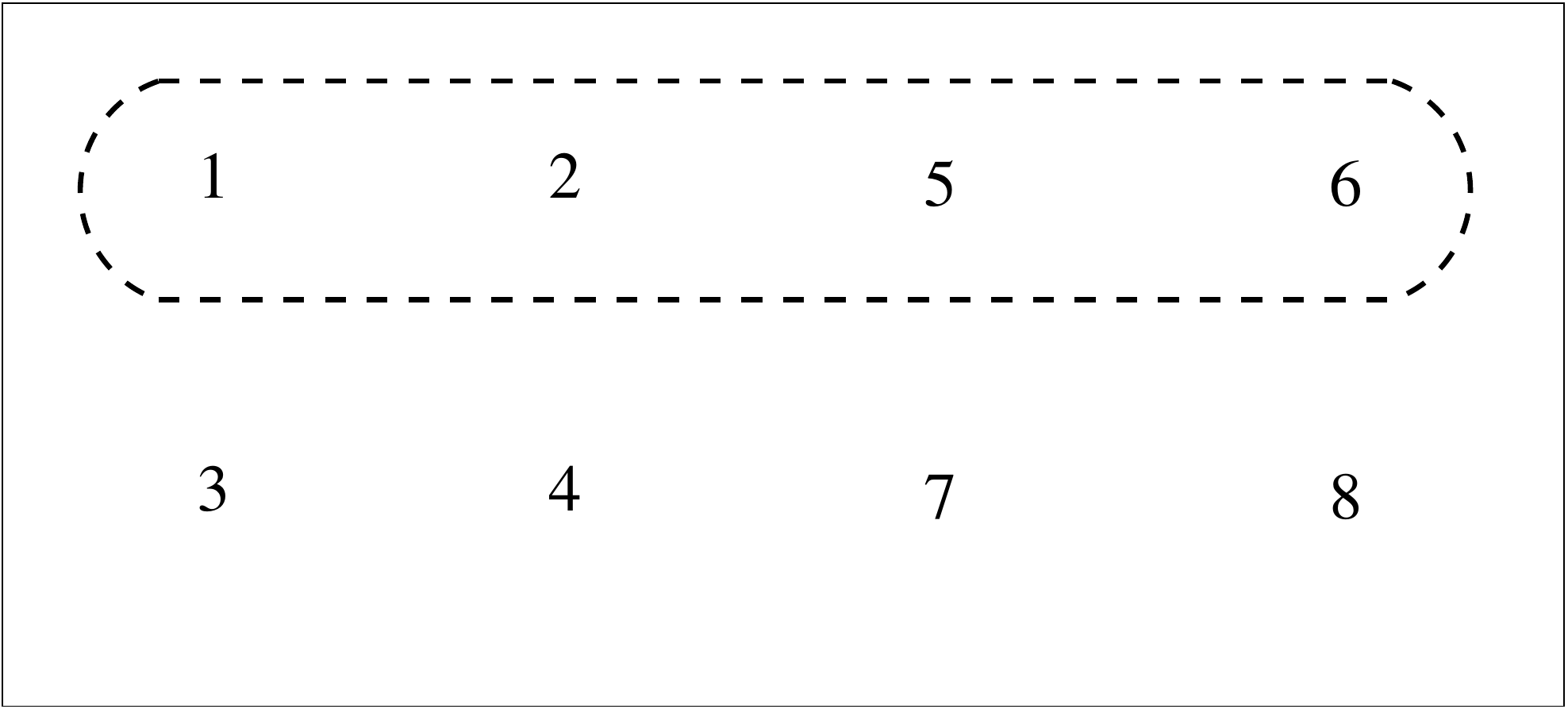}
\caption{Joint parity measurement $Z_a Z_b = \gamma_1 \gamma_2 \gamma_5 \gamma_6$.}
\label{fig:pinch}
\end{figure}

If the loop encircles a single hole (as in Ref.~\onlinecite{basismeasure}) or a pair of dislocations, then in fact one should just measure all qubits in the
$Z$ basis {\it inside} the loop.  Suppose the loop instead encircles, for example, $4$ dislocations from two different logical qubits in a dislocation code,
and the goal is to measure a joint parity $Z_a Z_b$ on both qubits.  In this case, to avoid measuring $Z_a$ and $Z_b$ separately, one must measure just in an annulus near the loop.
See Fig.~\ref{fig:pinch}, where logical $Z_a = \gamma_1 \gamma_2$ and logical $Z_b = \gamma_5 \gamma_6$.  If we measure all qubits inside the dashed line in the appropriate basis, we will measure both $Z_a$ and $Z_b$ separately; if the goal is to measure only $Z_a Z_b$, then we must restrict to measurements inside an annulus.

This technique allows us to perform a logical CNOT in a time that is $O(1)$, independent of the distance between the dislocations.  
This procedure of using joint parity measurements instead of braiding is reminiscent of the idea of measurement-only
topological quantum computing\cite{motqc}. 

The above measurement technique is used in the teleportation scheme of Ref.~\onlinecite{topt} to allow teleportation in time $O(1)$.
A second technique for fault tolerant measurement is as in Fig.~15 of Ref.~\onlinecite{measureturnoff}.  This involves changing stabilizers during the measurement.
This measurement technique is used to reduce the spacetime volume required to performed measurements.

\begin{figure}
\includegraphics[width=2in]{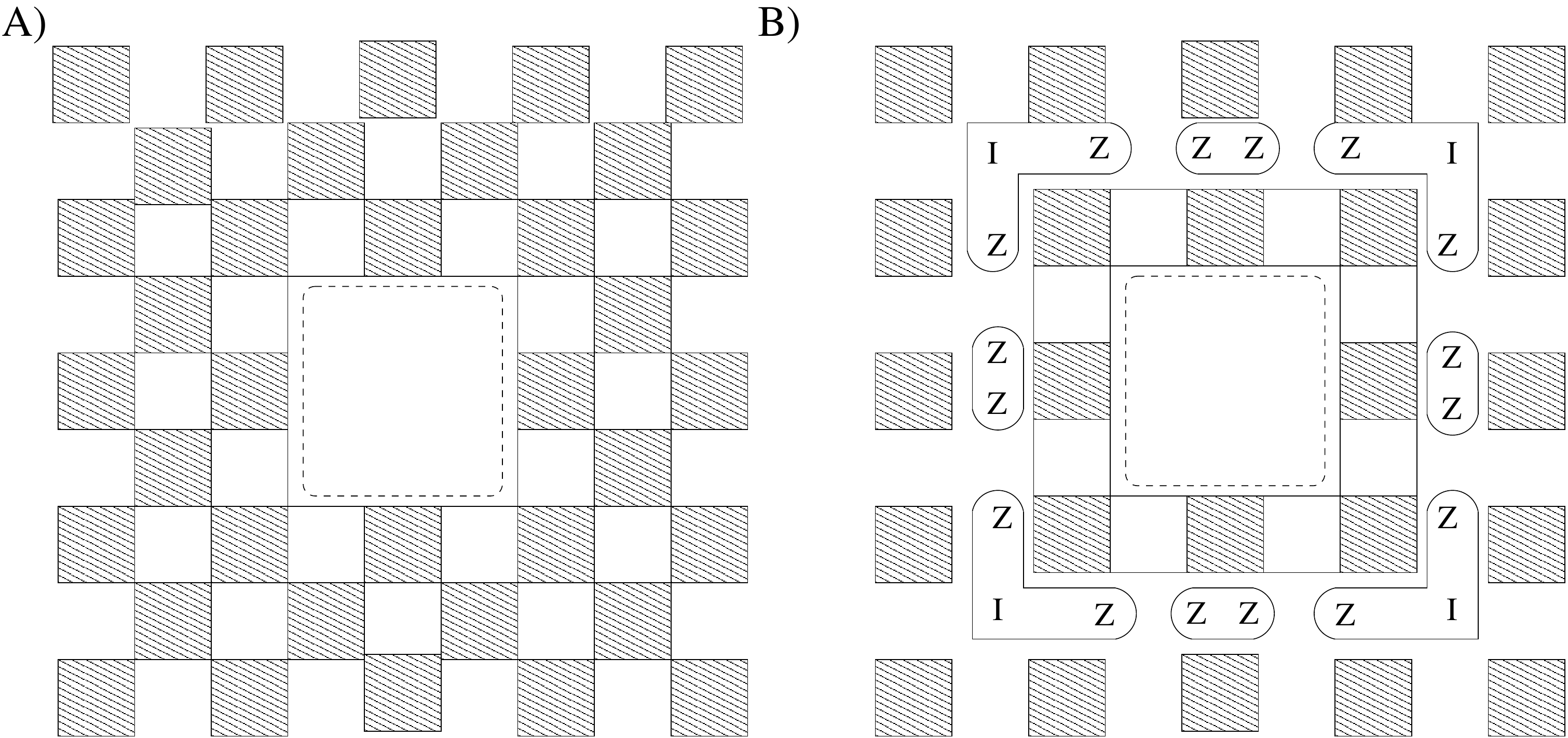}
\caption{Dashed line indicates some location containing dislocation pairs; interior geometry is arbitrary.  (a) Original lattice.  (b) Modified stabilizers.  Stabilizers are turned off in a ring, and additional stabilizers are turned on; we only indicate the
new stabilizers turn on which act on the outer boundary of the inner piece and additionally we do not show four stabilizers acting on the four qubits on the corners of the inner piece as in Fig.~\ref{fig:square}.}
\label{fig:hole}
\end{figure}

We now describe a third fault tolerant measurement that may be useful for measuring logical operators in a dislocation code (or indeed in any other code where we wish to measure a logical operator which is a product of Paulis around a loop).
Consider some location in a lattice, around which we wish to measure a logical operator.  This location may contain some even number of dislocations, and the particular geometry inside this region is not important for this construction.  We choose a gauge such that this operator is product of $Z$s, for example, around a loop.
Refer to Fig.~\ref{fig:hole}.
The original lattice is in (a), with the lattice extending further outside the area shown.  The dashed line indicates a location with some unspecified geometry inside.
We wish to measure a product of $Z$ in a loop containing this region.  We do this by changing stabilizers, turning off the stabilizers in a ring of plaquettes in (a); for example, we will turn them off in the ring just inside the outermost ring shown in (a).  This then disconnects the code into two pieces, as shown in (b).  We call these the inner and outer pieces; the inner piece contains the dashed line.  The $Z$s inside ovals in (b) denote additional stabilizers which we turn on involving qubits on the outer boundary of the inner piece.  The new stabilizers are the same as those in Fig.~\ref{fig:square}.  This choice of stabilizers prevents there from being any operators supported on the boundary which commute with the stabilizers but which are not themselves products of stabilizers.
We similarly turn on stabilizers which are products of $Z$s on the inner boundary of the outer piece; these stabilizers are not shown.

Crucially, the product of the added stabilizers around the outer boundary of the inner piece is equal to the desired logical operator.  Hence, the measurement of these
stabilizers gives the logical operator.  We cannot expect, of course, that these stabilizers will be measured perfectly.  However, we can determine the product using error correction.  We now sketch how to use
 minimum weight matching to perform this error correction. Crucially, to use minimum weight matching, we will need to run the minimum weight algorithm {\it twice}, to determine whether it is more likely that the product if $+1$ or $-1$.  We first run the algorithm under the assumption that the product is $+1$.
  The values of these boundary stabilizers are random (subject to the constraint on their product) after they are turned on; this can be mimicked by
assuming that the value is equal to $+1$ for all of them, but allowing there to be an error which flips the sign of a pair of neighboring stabilizers which occurs with probability $1/2$; i.e., by allowing such errors with $0$ weight to occur starting from the configuration where all are $+1$, the result is that all configurations with product $+1$ are equally likely.  Thus, the weight to match {\it any} pair of boundary stabilizers is equal to $0$.
We then repeat the matching assuming the product is minus $1$, which can be done by assuming that the value is $+1$ for all of them except for one
stabilizer where the value is $-1$, and repeat the same matching.
This gives two different choices, corresponding to whether the product of the stabilizers is $+1$ or $-1$; identifying the choice with the lowest weight determines what the
product of stabilizers is.  This matching can be carried out over some number of rounds, matching errors in spacetime.

In fact, the same matching can also be applied to the stabilizers on the inner boundary of the outer piece giving an additional way to infer the desired logical measurement.  Further, the product of
a stabilizer $ZZ$ on the outer boundary of the inner piece and another stabilizer $ZZ$ on the inner boundary of the outer piece 
is
a product of four $Z$s around a plaquette and the initial value of this product is known with some confidence (as that plaquette stabilizer would have been measured in the configuration in (a) in the figure).
Using all this information, however, goes beyond a matching algorithm.

\subsection{Numerical Results}
We have numerically analyzed the performance of the dislocation code in the simplest possible error model, assuming that we prepare an initial state, apply random noise, and then attempt to error correct using perfect measurement of the stabilizers.  A more realistic treatment, including errors in stabilizer measurement and even more a detailed study of circuits, is left for the future.  The noise model considered was that errors are produced with probability $p$ on each qubit, and if there is an error it is equally likely to be an $X$, $Y$, or $Z$ error.

We considered two different decoders.  One is the standard minimum-weight perfect matching decoder.  The second decoder is a greedy decoder, similar to that in Ref.~\onlinecite{greedy}.
This greedy decoder finds a pair of defect plaquettes with the minimum distance, and matches them; then it finds another pair with minimum weight among those remaining, and matches those, and so on.  The main difference from Ref.~\onlinecite{greedy} is in our treatment of edges.  When applying this decoder to a patch with open boundary conditions, where it is possible to match a defect in the bulk to an edge, if such a match is possible with weight $w$, we treat that as being equivalent in cost to matching two defects in the bulk with weight $2w$.  The reason for this choice is that in this way, it is equally costly to match two bulk defects to the edge each with weight $w$ or match them to each other with weight $2w$.

Various comparisons are made using patches (i.e., squares with alternating electric and magnetic boundary conditions) and dislocations. In the figures, ``square" refers to a choice of a patch using the geometry in this paper, while ``planar" refers to a geometry rotated $45$ degrees as in Fig.~3 of Ref.~\onlinecite{measureturnoff} which has a larger number of physical qubits for the given distance (i.e., the patch is still planar, but it is not square to the lattice directions but rotated).  For the dislocations code analysis, we took four dislocations with toroidal boundary conditions; it would be better for analysis of large codes to take a larger number and we leave this for the future.

\begin{figure}
\includegraphics[width=3in]{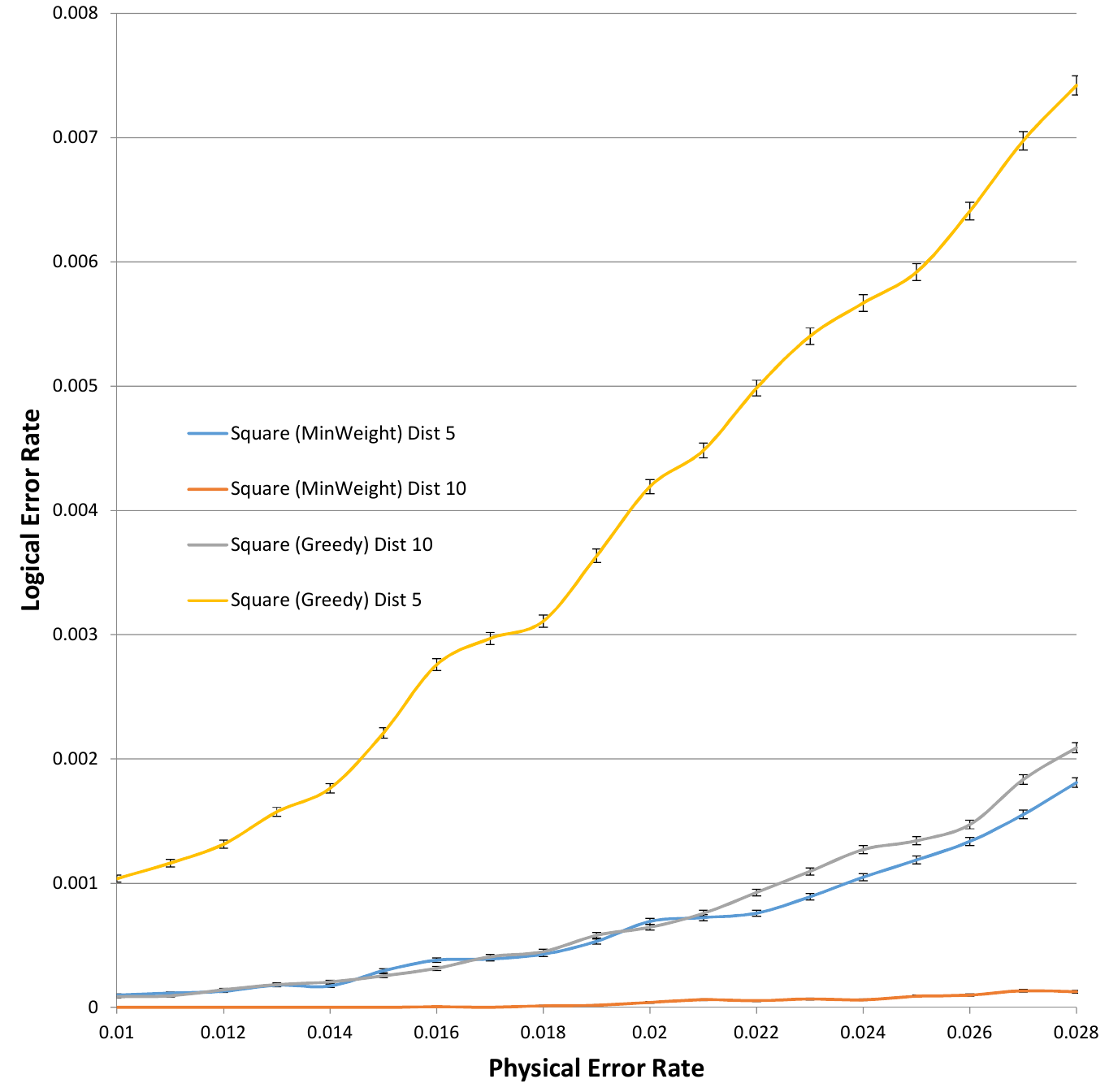}
\caption{Logical error rate as a function of physical error rate for square patches at low noise levels.}
\label{fig:low}
\end{figure}

In general, the minimum weight decoder is superior; see Fig.~\ref{fig:low} for an analysis at low noise levels.  The distance for the patch refers to the linear size of the patch, which in this case is the same as the code distance.  In fact, one reason for the difference is that the minimum weight decoder will always correct $k$ errors if $2k<d$ where $d$ is the code distance.  This is not, however, true for the greedy decoder, which can fail in this case.  Thus, the true asymptotic performance of the decoders is different at very low noise levels.  For example, for a distance $3$ code, both the  minimum weight decoder and correct all cases with $1$ error ($Z,X$, or $Y$ error).  However, for a distance $5$, the minimum weight decoder corrects all cases with $2$ errors, while the greedy decoder can fail on some cases with $2$ errors, although it does correct all cases with $1$ error.  The failures with $2$ errors for the greedy decoder at distance $5$ are rare, though; the probability that it fails for $2$ randomly chosen errors is $0.034...$ (we have found this by exactly enumerating all cases with $2$ errors).  Similarly, the failure probability with the greedy decoder for $3$ errors is higher at distance $5$ than it is for minimum weight; the probability that $3$ errors cause a failure for the minimum weight decoder is $0.037...$ while for the greedy decoder it is $0.11...$  At distance $7$, the minimum weight decoder can fix all cases of $1,2,3$ errors, but the greedy decoder can fail (albeit with probability $<0.002$) with only two errors.

\begin{figure}
\includegraphics[width=3in]{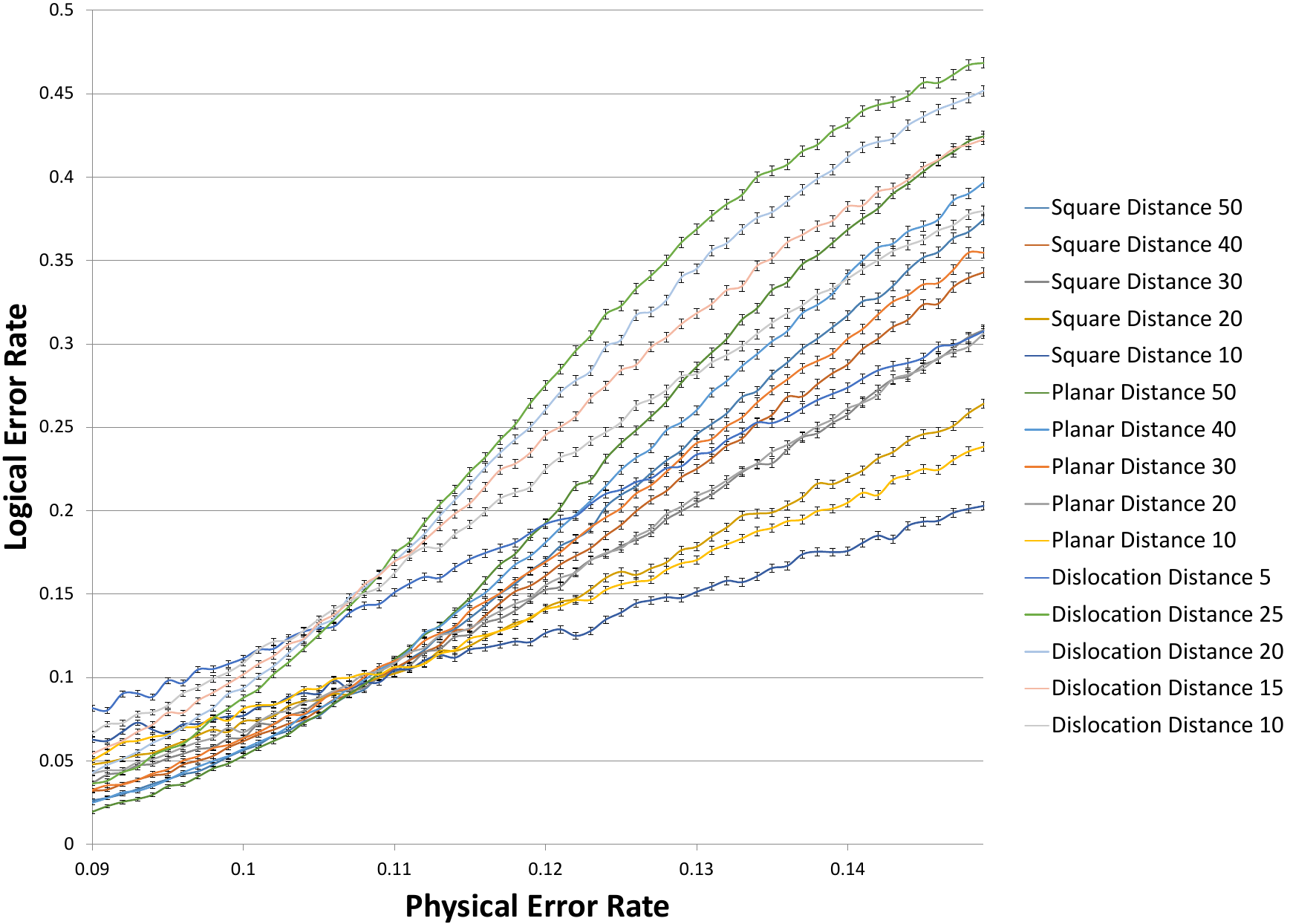}
\caption{Logical error rate as a function of physical error rate for greedy decoder.  The distance of the square and planar patches refers to the code distance.  The distance for dislocation refers to the distance between dislocations which is {\it half} the code distance.}
\label{fig:greedymaster}
\end{figure}

\begin{figure}
\includegraphics[width=3in]{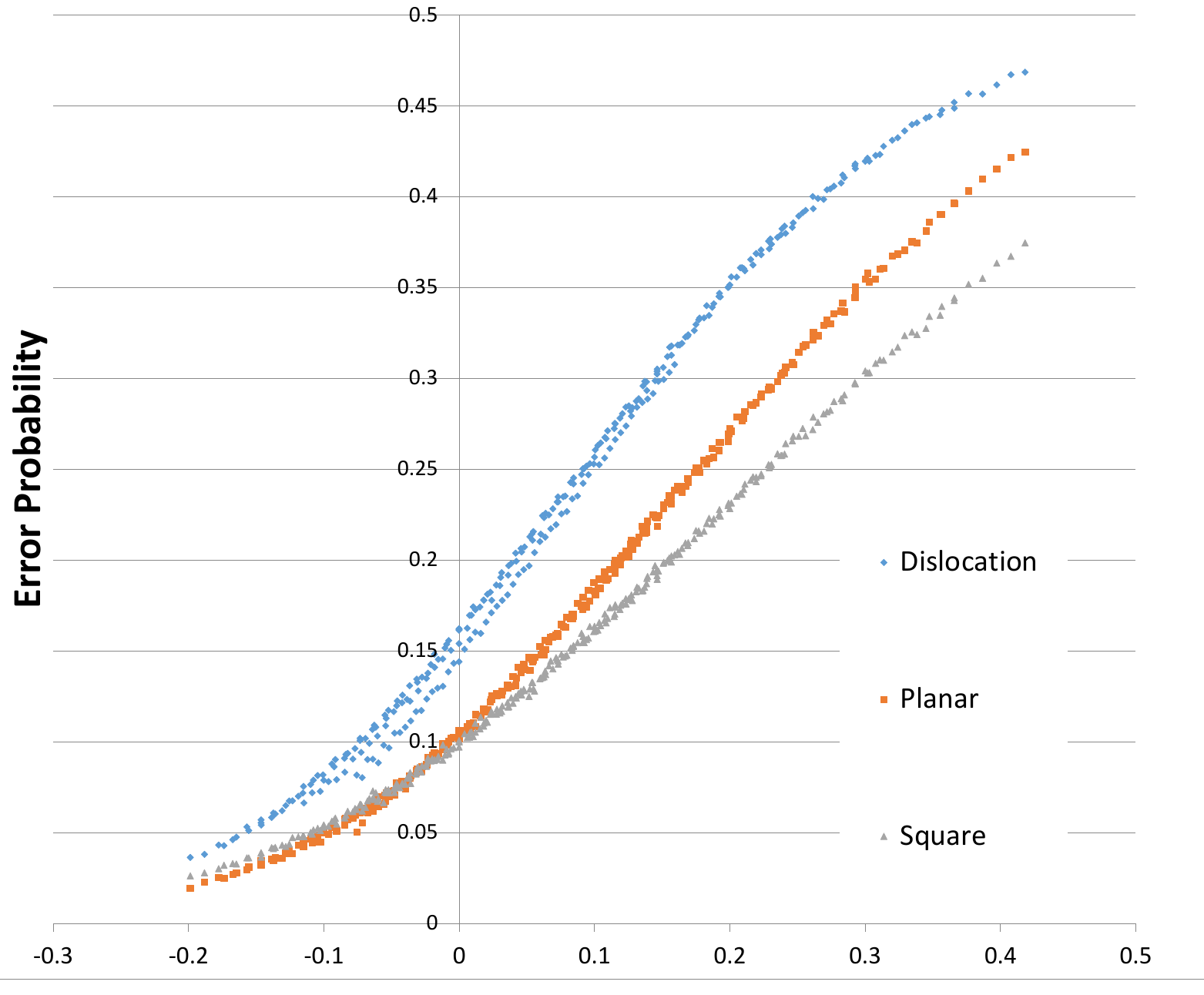}
\caption{Scaling collapse for greedy decoder.}
\label{fig:greedyscaling}
\end{figure}

The performance of the greedy decoder at larger noise is also worse, and its threshold seems to be around $p=0.109$.  See Fig.~\ref{fig:greedymaster} for this case for several geometries.
The threshold seems to be the same for both types of defects, as expected.
In Fig.~\ref{fig:greedyscaling} we show a finite-size scaling collapse, plotting the logical error probability as a function of $(p-p_c)*L^{\theta}$ with $p_c=0.109$ and $\theta=0.6$.  All curves for all different sizes for a given geometry are plotted with the same symbol.

\begin{figure}
\includegraphics[width=3in]{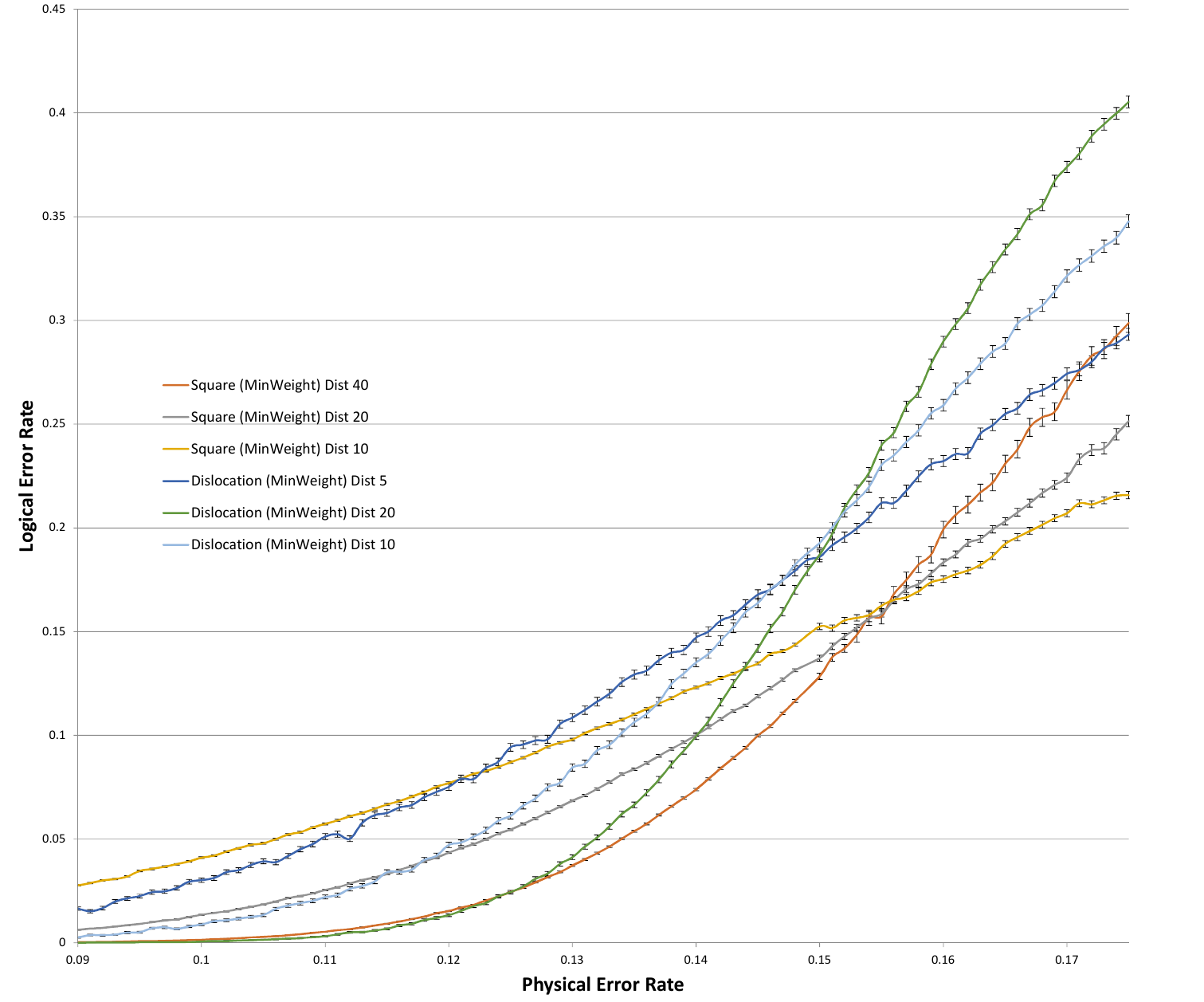}
\caption{Logical error rate as a function of physical error rate for minimum-weight decoder.  The distance of the square patches refers to the code distance.  The distance for dislocation refers to the distance between dislocations which is {\it half} the code distance.}
\label{fig:minweightmaster}
\end{figure}

\begin{figure}
\includegraphics[width=3in]{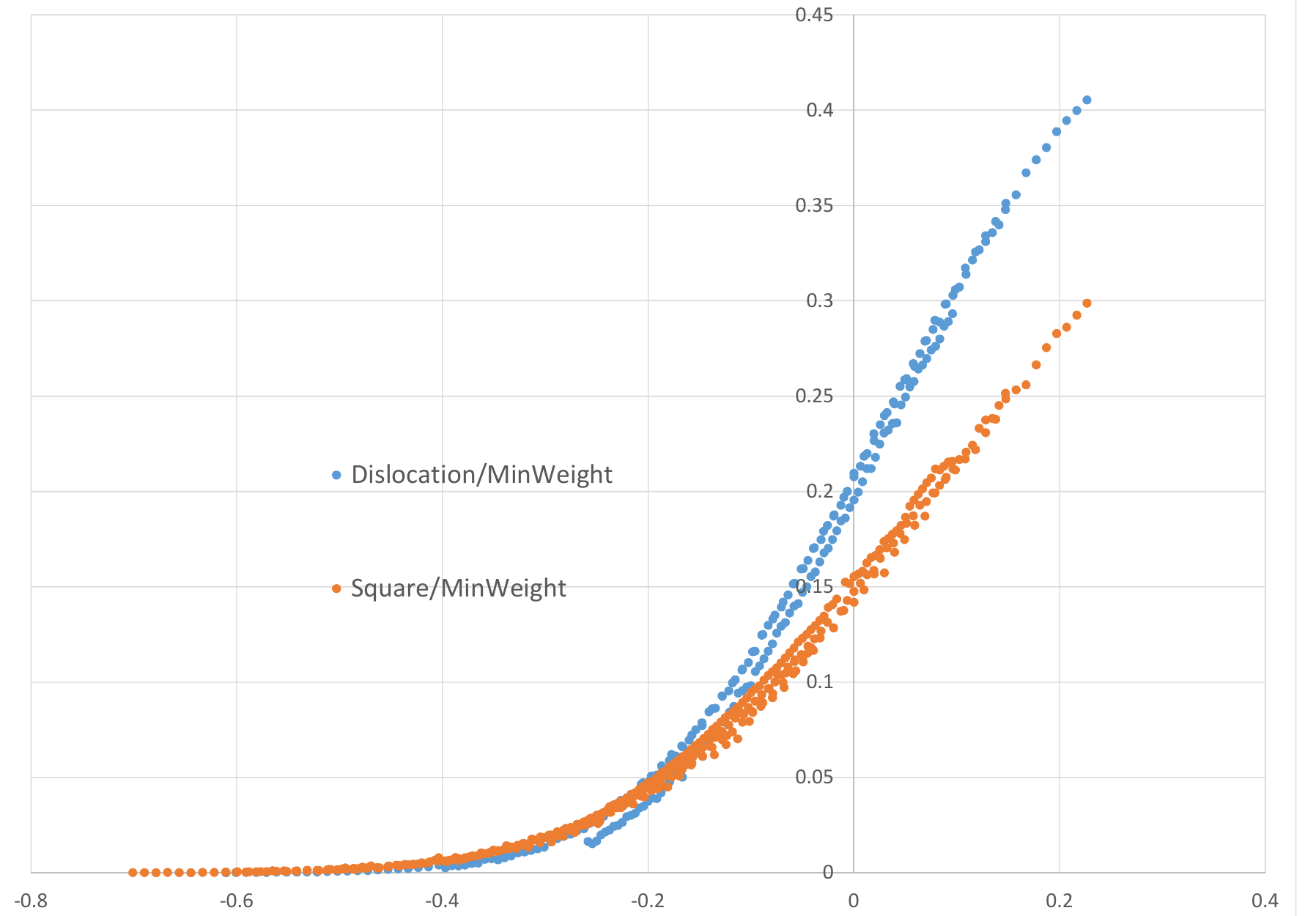}
\caption{Scaling collapse for minimum weight decoder.}
\label{fig:minscalecollapse}
\end{figure}

Finally, we show the performance of the minimum-weight perfect matching decoder in Fig.~\ref{fig:minweightmaster}.  Some difference between the threshold for dislocation and square geometries appears on this graph.  We believe that this is a finite size effect.  In Fig.~\ref{fig:minscalecollapse} we show a scaling collapse with $p_c=0.152$ for both square and dislocation geometries.  In fact, a slightly better fit is found for $p_c=0.15$ for dislocation and $p_c=0.155$ for square.

\section{Non-Clifford Operations}
The schemes of Refs.~\onlinecite{magic,topt} rely on using magic states to perform $T$ gates.
Using these $T$ gates that are produced, it is possible to perform universal quantum computation.  For definiteness, assume that we wish to perform a single qubit rotation $R(\theta)=\exp(-i \sigma^y \theta)$ by angle $\theta$ about the $Y$ axis (very frequently, quantum circuits are written in terms of Clifford operations and arbitrary angle single qubit rotations and we assume this when discussing online cost in the next paragraph).  Using $T$ gates and Clifford gates, we can approximate such rotations.
However, a logarithmic overhead in the accuracy of the desired rotations arises: to represent a  single-qubit rotation by an arbitrary angle $\theta$ to an accuracy $\delta$ will require logarithmically many (in $1/\delta$) rounds of $T$ gates and Cliffords.

However, the use of state injection need not be limited to implementing $T$ gates.  Here, we show that we can make the online cost {\it independent} of the desired accuracy $\delta$ by injecting arbitrary angles.
Here, by the online cost, we assume that some ancilla factory separately is preparing states of the form $Y(\theta)=\cos(\theta/2)|0\rangle+\sin(\theta/2)|1\rangle$ for a variety of desired angles, and our goal is to minimize the depth of the circuit that uses these ancillas.  One reason to consider this online cost is that
the ancilla factory can produce many ancillas in parallel, and so we may wish to cost the gates used to generate the ancillas separately.
The single qubit operations will be by angles $\theta$ drawn from some set of angles $\{\theta_1,\theta_2,...\}$.
Assume that we want to perform a rotation by each angle $\theta_i$ a total of $n_i$ times.
We will prepare in advance $n_i$ copies of $Y(\theta_i)$ for each $i$.  We will also prepare some number (explained below) of copies of states $Y(2\theta_i),Y(4\theta_i),...$

Then, when we wish to perform $R(\theta_i)$, we perform state injection using a copy of $Y(\theta_i)$.  If successful, this performs $R(\theta_i)$.
If unsuccessful, it performs $R(-\theta_i)$.  In the later case, we immediately follow with a state injection of $Y(2\theta_i)$.  If successful, we perform $R(2\theta_i) R(-\theta_i)=R(\theta_i)$.  If again unsuccessful, we perform state injection with $Y(4\theta_i)$, continuing in this way until we
are successful.  See Refs.~\onlinecite{ancilla}; however, those references focused on the case of a single qubit and did not consider the parallelization issue which we now address.

Naively, this scheme still leads to a logarithmic overhead in time: each attempt to perform the desired rotation gate has a probability $1/2$ of succeeeding.  If we have $N$ qubits and need to perform rotation gates on all of them, it will take a time roughly $\log_2(N)$ until all of the rotation gates succeed.
However, there is no need for other qubits to wait until all of the rotations are done.  Suppose, for example, that $N=4$, and on the first round of the circuit we apply single qubit rotations to all $4$ qubits; on the second round,  we wish to apply a CNOT gate from qubit $1$ to qubit $2$ and another CNOT gate from qubit $3$ to qubit $4$,
and then on the third round we again wish to apply single qubit rotations on all for qubits.
Suppose further that the single qubit rotations succeed on qubits $1,2,3$ on the first attempt on the first round of the circuit, but fail on qubit $4$.  In this case, we can perform the CNOT gate from qubit $1$ to qubit $2$ without waiting for qubit $4$ to succeed, while qubit $3$ must wait idle until qubit $4$ finishes.  Then, qubits $1,2$ can start their single qubit rotations on round $3$ of the circuit immediately after doing the CNOT, without waiting.
Thus, the parallelization question arises: if we have a circuit with gates, and a gate can execute once all of its input gates finish, and if a gate has a certain probability of finishing each time we try it, what is the overall slowdown?  We define this scheme more formally in the next subsection and we show that given a quantum circuit with $r_{tot}$ rounds of gates and $N$ qubits, the probability that it takes a time $T$ to finish decays exponentially in $T-T_0$ for $T>T_0$, where $T_0={\rm const.}*r_{tot}+{\rm const.}*\log(N)$.  Finally in subsection \ref{req}, we discuss the number of ancillas required and certain implementation details.

One important aspect of the parallelization scheme is that we will have to assume that all gates in the circuit have a bounded number of wires, and the results depend upon this bound.  That is, it is advantageous to have a circuit which does not use completely arbitrary Cliffords but rather to use those Cliffords which can be decomposed into a product of gates acting only on a few qubits at a time.  To understand this, consider the example above and suppose that instead of performing a CNOT from qubit 1 to 2 and a CNOT from qubit 3 to 4 on the second round of the circuit, we desired to perform some general Clifford operation on all four qubits which could not be decomposed as a product of operations on less than four qubits; in this case case, the Clifford would have to wait until {\it all} qubits finished the first round.
This would be a potential advantage of using the
circuits in Ref.~\onlinecite{improving}, for example, in quantum chemistry applications which use fewer Clifford gates; that is, while it is sometimes supposed that there is little advantage in reducing the number of Clifford gates since the non-Clifford gates dominate the costs, in any kind of scheme where the success of the non-Clifford gates is probabilistic, there may be an advantage to simplifying the Clifford gates as it may allow one to start some of the non-Clifford gates earlier.

\subsection{Parallel Scheme}
We now define a formal setting for the parallelization issue raised above.  This setting in fact has nothing to do with quantum mechanics; it would be applicable to any situation
in which computation is done in a circuit, where some circuit elements take a time to finish that is drawn from an exponential distribution, and a given element cannot start computing
until all of its inputs have finished.

We consider a circuit diagram with gates connected by wires.  There will be $N$ incoming wires at the start of the circuit and $N$ outgoing wires at the end.
The gates are organized into ``rounds".  Inputs to gates may be either incoming wires or outputs of other gates.
Gates whose inputs consist solely of incoming wires to the circuit will be on round $0$; otherwise, the round of a gate $G$ is equal to $1$ plus the
maximum round of gates $G'$ such that an output wire of $G'$ is an input to $G$.
Let there be a total of $r_{tot}$ rounds.
Each gate has some number of incoming wires, where the number of incoming wires is bounded by some constant $D$.  
We assume that each gate has the same number of incoming as outgoing wires (one could perhaps consider generalizing to the case that this is not true; note however that if the total number of wires entering or leaving all gates in a given round is bounded and the number of incoming and outgoing wires on every gate is bounded then we can add some ``dummy wires" to return to the case where each gate has the same number of incoming and outgoing wires).

Time will proceed discretely and will be labelled by an integer.  The evolution will start at time $1$.  We will define a discrete Markov process, which will label each wire in the circuit by some integer.  It will also label each gate by a time at which that gate ``finishes".  We start by labelling all the incoming wires to the circuit diagram by the integer $0$.  Initially, no gates have ``finished" and so all gates are unlabelled.  At time $t$, let ${\cal P}(t)$ denote the
set of gates which have not yet finished and for which all incoming wires are labelled by a time less than $t$.
For each gate in ${\cal P}(t)$,
with probability $P$, we label the gate as finishing at time $t$ and we also label the wires leaving the gate with time $t$; otherwise, with probability $1-P$, no change is made to that gate and those wires.  These choices are made independently for each gate in ${\cal P}(t)$.
The circuit is considered to finish when all gates have finished.

We will assume that all gates have the same probability $P$ of finishing; this
is not true for the quantum application above, as the Clifford gates always succeed.  However, assuming a probability $P$ of finishing for all gates
gives a more pessimistic estimate than assuming a probability $P$ of finishing for some gates and a probability $1$ of finishing for others.

Our main result, proven in the next subsection is:
\begin{theorem}
Given any $D$ and any $P>0$, there exist constants $c_1,c_2$ and $c_3>0$ such that the following holds.  For any such circuit with $N$ incoming and outgoing wires on each round (this $N$ is the number of
qubits) and $r_{tot}$ rounds, the probability that the circuit has not finished in a time $T$ is bounded by
$\exp(-c_3 (T-T_0))$, where
\be
T_0=c_1 r_{tot} + c_2 \log(N).
\ee
\end{theorem}

In fact, our proof will allow adversarial adjustment of the circuit.  In particular, consider any sequence of events up to time $t$.  Then, there will be some set $W$ of wires which are already labelled but which enter gates which have not yet finished.  We allow the adversary to change arbitrarily the gates which have not yet finished, subject to the constraints that each gate have the same number of incoming and outgoing wires, with at most $D$ such wires.

\subsection{Amortized Analysis of Parallel Scheme}
We now perform an amortized analysis of the parallel scheme.
For analysis purposes, we modify the circuit as follows.  Given any circuit, we add additional gates with at most $D$ incoming wires on rounds $r_{tot}+1,r_{tot}+2,...$, extending the circuit indefinitely.
These gates may be thought of as identity gates so that they simply preserve the data without doing any further computation.  Then, the original circuit finishes when all gates in the first $r_{tot}$ rounds finish in the modified circuit.  This is done to simplify the analysis so that we do not need to separately handle the time at which the circuit finishes.  When we refer to the computation ``finishing" later, we mean that all gates in the first $r_{tot}$ rounds finish.

Also, to simplify the analysis, we add additional identity gates to the circuit as follows.  Suppose that a wire leaves a gate $G$ on round $r$ and enters a gate $G'$ on some round $r'>r+1$.  In this case, we modify the circuit by adding additional identity gates $G_1,G_2,...$ on rounds $r+1,r+2,...,r'-1$ with $1$ incoming and $1$ outgoing wire each, and connect the wire leaving $G$ into gate $G_1$, then connect the output of $G_1$ into $G_2$, and so on, and finally into $G'$.  In this way, there will always be a total of $N$ incoming wires to each round.

We define a {\it weight} function after $r$ rounds of the circuit.  Let $n(t,r)$ be the number of wires that leave gates at round $r$ which finish by time $t$ (these gates may possibly finish at some earlier time) such that these wires enter gates at round $r+1$ which have {\it not} finished by time $t$; i.e., let $S(t,r)$ be the set of wires $w$, such that $w$ is labelled by a time at most $t$ and $w$ enters a gate that has not finished by time $t$ and let $n(t,r)=|S(t,r)|$.
Let 
\be
\label{Ctrdef}
C(t,r)=\sum_{r' \leq r}n(t,r).
\ee

We can explain the definition of $n(t,r)$ differently using an equivalent definition of the Markov process defining this model.  Note that the circuit can be implemented using $N$ qubits.  At every time, we have $N$ qubits, and each qubit is labelled by some wire in the circuit.  Initially, each qubit is labelled by a different incoming wire.  Then, if a gate finishes at time $t$, we remove the qubits labelled by the incoming wires of the qubit and replace them with qubits labelled by the outgoing wires of the circuit.  Then, $n(t,r)$ counts the number of qubits at time $t$ labelled by an outgoing wire of a gate on round $r$, while $C(t,r)$ counts the number of qubits labelled by an outgoing wire of a gate on round at most $r$.  Note that $C(t+1,r) \leq C(t,r)$.

Define a function $W(t,r)$ by
\be
W(t,r)=r-\frac{1}{A}\ln(C(t,r)),
\ee
where $A$ is a constant to be optimized later.
Define the {\it weight} after time $t$, $W(t)$, by
\be
W(t)={\rm min}_{r} W(t,r),
\ee
where the minimum is over all $r$ for which $C(r,t)$ is non-zero.

Define
$r_{last}(t)$ to be the minimum $r$ such that $n(t,r)>0$.
Note that
\be
W(t) \leq r_{last}(t),
\ee
and hence if the computation is not finished then $W(t)\leq r_{tot}$.

Our analysis is based on showing that the average value of $W(t+1)$
is at least equal to $W(t)$ plus some positive constant computed below; further, we will show that $W(t)\geq W(0)+vt$ for some $v>0$ with probability
that is exponentially close (in $t$) to $1$; see lemma \ref{explemma}.
This implies the theorem since $W(0)=-A^{-1} \log(N)$ and since if $W(t)>r_{tot}$ then the computation finishes.
To obtain the constants in the theorem, put $c_1=1/v$, $c_2=1/(Av)$.

We briefly motivate the choice of the weight function as follows (this paragraph is purely heuristic and does not play any role in the proof).  Suppose at some time $t$ we have some given $n(t,r)$.  Suppose all the gates are single wire gates so that
$D=1$.  Then, roughly $(1-P) n(t,r_{last}(t))$ of the gates in round $r_{last}(t)$ will not have finished at time $t+1$; roughly $(1-P)^2 n(t,r_{last}(t))$ will not have finished at time $t+2$, and so on.  We expect that eventually at a time
roughly $t+\log_{1/(1-P)}(n(t,r_{last}(t)))$ the last such gate will finish.  Thus, the computation might get delayed by an amount roughly $t-r_{last}(t)+\log_{1/(1-P)}(n(t,r_{last}(t)))$ at time $t+\log_{1/(1-P)}(N)$, where the {\it delay} at time $t$ is equal to $t-r_{last}(t)$.  However, we can also apply the same analysis to the set of $C(t,r_{last}(t)+i)$ gates on rounds $r_{last}(t),...,r_{last}(t)+i$
which finished at time $t$; at time roughly $t'=t+\log_{1/(1-P)}(C(t,r_{last}(t)+i))$ we expect that the last of these will have completed at least one more round, and hence at that time we will have $r_{last}(t')$ at most equal to $r_{last}(t)+1$ (it may of course be less if it is one of the gates from rounds less than $r_{last}(t)+i$ that have not completed).  The function $W(t)$ we have defined is a minimum over all choice of $r$ of a quantity inspired by this delay calculation: roughly, it is $t$ minus an estimate of the delay.  The reason for the constant $A$ is for technically optimizing estimates later.

One further reason for our definition that $W(t,r)=r-\frac{1}{A}\ln(C(t,r))$, rather than $W(t,r)=r-\frac{1}{A}\ln(n(t,r))$, is that given
our definition of $W(t,r)$, we have the property that $W(t+1,r)\geq W(t,r)$ always.

We now prove the following lemma:
\begin{lemma}
\label{explemma}
Given any $D$ and any $P>0$, there exist constants $v>0$ and $c'>0$ such that the probability that $W(t)\leq vt+W(0)$ is bounded by
\be
\label{markov}
{\rm Pr}[W(t) \leq W(0)+vt] \leq \exp(-c' t).
\ee
\begin{proof}
Consider some given situation after time $t$; i.e., our analysis is for a given situation of events on previous times.
We will first estimate the average increase $\overline{W(t+1)}-W(t)$, where the overline denotes the averaging over possible events at time $t+1$.
Consider a given $r$.  Let $S(t,r)$ be the set defined above Eq.~(\ref{Ctrdef}).  On round $r+1$, each of these
wires must participate in either a one wire gate or a multi-wire gate.  If it participates in a one wire gate, then it has a probability $P$ of finishing
round $r+1$ at time $t+1$.  If it participates in a multi-wire gate, it is possible that it must wait for some other gate in round $r-1$ to finish if the other wire in the gate is not in $S(t,r)$; note that
the other wires in the gate cannot be in $S(t,s)$ for $s>r$ (the addition of identity gates described above prevents this case).
If $n(t,r) \leq (D-1) C(t,r-1)$, it is possible that every single wire in $S(t,r)$ must wait for some other gate to finish round $r-1$.
However, if $n(t,r)>(D-1) C(t,r-1)$, then there must be some gates on round $r+1$ which do not need to wait.  Indeed, the number of wires entering gates in round $r+1$ which do
not need to wait is at least equal to
\be
K(t,r) \equiv n(t,r)-(D-1) C(t,r-1).
\ee

If $W(t,r)>W(t)+1/2$, then we say that round $r$ is {\it not important}.  We will show that, for sufficiently small $A$, the quantity $W(t,r)$ is likely to increase (and we estimate how likely it is to increase) by at least a constant for the rounds $r$ which are important.  We do not consider the change in $W(t,r)$ for the rounds which
are not important as they will have little effect on the minimum over $r$.

Assume then that $r$ is important.  Then $C(t,r)\geq \exp(A/2) C(t,r-1)$.  Note that $C(t,r)-C(t,r-1)=n(t,r)$.
Hence,
\be
n(t,r) \geq \Bigl( \exp(A/2)-1 \Bigr) C(t,r-1),
\ee
and so
\begin{eqnarray}
\frac{K(t,r)}{C(t,r)} &=&\frac{n(t,r)-(D-1)C(t,r-1)}{n(t,r)+C(t,r-1)} 
\\ \nonumber
&\geq &\frac{\exp(A/2)-D}{\exp(A/2)} \\ \nonumber
&=& 1-D\exp(-A/2) \\ \nonumber
&&\equiv \omega,
\end{eqnarray}
where the last line of the above equation serves as a definition of $\omega$.
On average at least $P K(t,r)$ of the wires in $S(t,r)$ enter gates which finish round $r+1$ at time $t+1$.
Hence, $C(t,r)-\overline{C(t+1,r)} \geq P K(t,r)$.
Using concavity of the logarithm,
\be
\overline{W(t+1,r)}\leq r-\frac{1}{A}\ln(C(t,r)-P \omega C(t,r))=W(t,r)+\frac{1}{A}\ln(1-P \omega).
\ee
For sufficiently large $A$ so that $\omega>0$, this means that on average $W(t+1,r)-W(t,r)$ is greater than some positive constant.

This does not yet give what we want; we want to show some lower bound on the probability
that $W(t+1,r)-W(t,r)$ is greater than some positive constant for all important rounds $r$.
However, by the assumption that $r$ is an important round, we have
$C(t,r)\geq\exp(A (r-r_{last}-1/2)$ so
\be
K(t,r) \geq \omega \exp(A (r-r_{last}-1/2)).
\ee
Hence, $K(t,r)$ is exponentially large in $r$.  The probability that less than $P K(t,r)/2$ wires in $S(t,r)$ enter gates which finish round $r+1$ at time $t+1$ is exponentially small in $K(t,r)$ as the probabilities that different gates finish are independent (the particular constant $P/2$ in $P K(t,r)/2$ is unimportant and any constant in $(0,P)$ would suffice).  Since this probability is exponentially small in $K(t,r)$ it is doubly exponentially small in $r$.
If at least $P K(t,r)/2$ wires do enter gates which finish round $r+1$ at time $t+1$ then $W(t+1,r)\geq W(t,r)+A \ln(1-P\frac{\omega}{2})$.
We sum over these probabilities and apply a union bound to upper bound the probability that there is an important round $r$ such that $W(t,r)$ does not increase by some strictly positive constant.  For sufficiently large $A$, we can bound this probability less than $1$.
Hence, for some sufficiently large $A$, there are some constants $p,c>0$ such that, with probability at least $p$, $W(t+1,r)-W(t,r)\geq c$ for all important rounds.
Hence, for sufficiently large $A$,
\be
W(t+1) \geq W(t)+{\rm min}(1/2,c) = W(t)+c',
\ee
with probability at least $p>0$, where $c'={\rm min}(1/2,c)$.

This already gives sufficient information to show that $\overline{W(t)}\geq W(0)+pc'$.  Note that $W(t+1)\geq W(t)$ always.
However, we can also show that it is exponentially unlikely (exponentially in $t$) for $W(t)$ not to be at least a constant times $t$ larger than
$W(0)$; the proof of this will be similar to the proof of the Chernoff bound.
Let $a$ be a negative constant to be chosen later.
Note that $\overline{\exp(a W(t+1))} \leq \Bigl((1-p)+p\exp(ac') \Bigr) \exp(a W(t))$.
Hence $\overline{\exp(a W(t))} \leq \Bigl((1-p)+p\exp(ac') \Bigr)^t \exp(a W(0))$.
Hence, by Markov's inequality, the probability that $W(t)\leq W(0)+vt$ is bounded by
\be
\label{mark2}
{\rm Pr}[W(t) \leq W(0)+vt] \leq \frac{\Bigl((1-p)+p\exp(ac') \Bigr)^t}{\exp(avt)}.
\ee
For any given
$v \leq pc'$, we can find an $a<0$ such that the expression above is exponentially small in $t$.
Hence the lemma follows.  To obtain the constants in the lemma, fix some definite $v<pc'$ and minimize the expression on the right-hand side
over choice of $a$ in Eq.~(\ref{mark2}).
\end{proof}
\end{lemma}

\subsection{Number and Accuracy of Ancillas Required, and Implementation Details}
\label{req}
Finally, we consider the number of copies of states $Y(2\theta_i),Y(4\theta_i),...$ that we will need.  We consider two different regimes, depending on the magnitude of $n_i$.  Suppose there are a total of $A$ different angles that we need in the entire circuit, indexed by $i=1,...,A$, and suppose that $n_i=1$ for all $i$.  In this case, for each $i$, we need roughly $\log_2(A)$ ancillas, with one copy each of $Y(\theta_i),Y(2\theta_i),...,Y(2^{\log_2(A)}\theta_i)$, in order for the entire computation to be likely to succeed.
This leads to an unfortunate logarithmic overhead in the number of ancillas that we need, which may be expensive as the ancillas $Y(\theta_i)$ already need to be prepared to high accuracy and hence are expensive (we discuss the accuracy needed in the ancillas in the next paragraph).  However, in many applications, we will have $n_i>>1$.  For example, in applications in quantum
chemistry using Trotter-Suzuki evolution, it may be necessary to have a large number $A>>1$ of angles (encoding the large number of coupling constants in the Hamiltonian), but each angle will be used many times, as there will be many different Trotter steps.  If $n_i = \Omega(\log^2(A))$ for all $i$, then there is indeed only a {\it constant} overhead in the number of ancillas required as may be seen as follows.  Pick some constant $c>1/2$.  It is exponentially unlikely (exponentially in $n_i$) that more than $cn_i$ of the gates $R(\theta_i)$ will fail on the first attempt.  Hence, if we prepare $cn_i$ copies of $Y(2\theta_i)$ for each $i$, then it is exponentially unlikely that we will not have enough copies of $Y(2\theta_i)$ for the given $i$.  Hence, if $n_i \gtrsim \log(A)$, it is unlikely that there will be any $i$ for which we will not have enough copies of $Y(2\theta_i)$.
Similarly, it is exponentially unlikely that $c^2n_i$ copies of $Y(4\theta_i)$ will not suffice, and in general $c^a n_i$ copies of $Y(2^a \theta_i)$ will suffice so long as $c^an_i$ is large compared to $\log(A)$.  Once we reach large enough $a$ that $c^an_i$ is of order $\log(A)$, this estimate breaks down, but then we know that $c^an_i \log(A) \sim \log^2(A)$ extra ancillas suffice (by the argument above in the regime that $n_i=1$).  By summing the geometric series $n_i,cn_i,c^2n_i,...$ up until $c^an_i$ and then adding $\log^2(A)$, we find indeed that there is at most a constant overhead.
This regime is quite relevant to the quantum chemistry simulation considered in Ref.~\onlinecite{coalesce}; further, that regime has the advantage that by
coalescing different terms by different magnitude we can change the angles needed, possibly reducing the number of different angles required\cite{wecker}.

Now we consider the accuracy of the ancillas that we need.  Suppose we need to prepare $Y(\theta_i)$ to an accuracy $\delta$ in order to implement the gate $R(\theta_i)$ to the desired accuracy.  One may worry that we will need to prepare the ancillas $Y(2\theta_i),Y(4\theta_i),...$ to higher and higher accuracy.  Suppose however that there are a total of $R$ different rotation gates in the circuit that we wish to implement.  It is common to argue that ancillas should be prepared to an accuracy $\delta\lesssim 1/R$ to ensure that the total  error will be small.  This estimate may be pessimistic as it assumes that errors add in the worst case rather than possibly averaging.  However, if we continue to use this estimate, then since the average number of ancillas that we use to implement the circuit is only a constant amount larger than the number of gates (in fact, twice as large since each gate has probability $1/2$ of succeeding), then we need to only increase the accuracy $\delta$ by a factor of $2$.  Further, the accuracy required depends slightly upon the type of errors that arise in state preparation.  Suppose that the dominant error is that rather than preparing a state $Y(\theta)$ with the desired angle $\theta$, we prepare the state $Y(\theta')$ for some other angle $\theta'$.  In that case, rather than trying to prepare the ancillas $Y(\theta),Y(2\theta),Y(4\theta)$, we prepare a sequence of ancillas $Y(\theta'),Y(\theta_2'),Y(\theta_3'),...$, where $\theta'$ is an approximation to $\theta$ within accuracy $\delta$, and
$\theta_2'$ is an approximation to $\theta+\theta'$ to within accuracy $\delta$, and in general $\theta_{k+1}'$ is an approximation to $\theta+\theta_2'+...+\theta_k'$ to within accuracy $\delta$.

In actual practice, using this scheme requires many teleportation steps.  For example, if the first state injection fails, one might teleport the qubit elsewhere and try a second state injection.  If that state injection also fails, the qubit must be further teleported elsewhere for yet another state injection.  This means that even after a state injection succeeds on the qubit, some time might be spent teleporting it back to the desired location.  However, this sequence of teleports to bring it back can be done in a time that at most doubles the time to do the state injection.  Imagine a sequence of regions ${\cal R}_1,{\cal R}_2,{\cal R}_3,...$; each region ${\cal R}_i$ shares two entangled pairs with ${\cal R}_{i+1}$.  One teleports the qubits from ${\cal R}_i$ to ${\cal R}_{i+1}$ and tries state injection.  If unsuccessful, one then teleports to ${\cal R}_{i+2}$ and again tries state injection, and so on.  After successful state injection, one uses the unused entangled pairs to teleport back in the reverse direction.

\section{Discussion}
We have analyzed codes based on dislocations, as originally suggested in Refs.~\onlinecite{dis2,dis}.  We have constructed circuits to perform stabilizer measurements and shown how to build logical operations.  A comparison to other surface codes shows that they
achieve higher density of logical qubits at the same distance.  One disadvantage in practical implementations using, for example, Josephson junction qubits, is that they will require an irregular layout.
Another potential disadvantage is the need for some stabilizers involving $5$ qubits as discussed above.

We have then analyzed a parallelization question relevant to injecting arbitrary rotations into a quantum circuit.  We have shown that this can be done, reducing time costs by a logarithmic factor in the desired precision, at the cost of a constant increase in spacetime volume required.

{\it Acknowledgments---} We thank G. Dauphinais, A. Fowler, D. Poulin, B. Smith, K. Svore, and D. Wecker for useful discussions.

\end{document}